\documentclass[12pt,draftcls,onecolumn]{ieeetran}
\usepackage{graphicx}
\usepackage{amsmath}
\usepackage{times}
\usepackage{subfigure}
\usepackage{cite}
\usepackage{authblk}
\usepackage{amssymb}
\usepackage{geometry}
\renewcommand{\baselinestretch}{1.5}

\def\diag{{\rm diag}\,}

\def\Exp{{\mathbb{E}}\,}
\def\tr{{\rm tr}\,}

\def\diag{{\rm diag}\,}
\def\real{{\rm Re}\,}
\def\imag{{\rm Im}\,}

\def\be{\begin{equation}}
\def\ee{\end{equation}}
\def\ba{\left[\begin{array}}
\def\ea{\end{array}\right]}
\def\bea{\begin{eqnarray}}
\def\eea{\end{eqnarray}}

\newcommand{\mb}[1]{\mathbf{#1}}
\newcommand{\mr}[1]{\mathrm{#1}}
\newcommand{\mc}[1]{\mathcal{#1}}
\newcommand{\ol}[1]{\overline{#1}}

\def\ba{{\bf a}}

\def\br{{\bf r}}

\def\bv{{\bf v}}

\def\bff{{\bf f}}
\def\bg{{\bf g}}

\def\o{\mathrm{o}}
\def\e{\mathrm{e}}

\newtheorem{theorem}{\textbf{Theorem}}

\newtheorem{lemma}{\textbf{Lemma}}

\begin{document}
\title{Interference Cancellation at the Relay for Multi-User Wireless Cooperative Networks
}
%
\date{}
\author[1]{Liangbin Li}
\author[2]{Yindi Jing}
\author[1]{Hamid Jafarkhani\thanks{Part of this work was presented at IEEE Wireless Communication \& Networking Conference (WCNC) 2010.}}
\affil[1]{{Center for Pervasive Communications \& Computing,

University of California, Irvine}} \affil[2]{University of Alberta}
%
%
\maketitle
\renewcommand{\baselinestretch}{1.6}
\begin{abstract}
We study multi-user transmission and detection schemes for a
multi-access relay network (MARN) with linear constraints at all
nodes. In a $(J, J_a, R_a, M)$ MARN, $J$ sources, each equipped with
$J_a$ antennas, communicate to one $M$-antenna destination through
one $R_a$-antenna relay. A new protocol called IC-Relay-TDMA is
proposed which takes two phases. During the first phase, symbols of
different sources are transmitted concurrently to the relay. At the
relay, interference cancellation (IC) techniques, previously
proposed for systems with direct transmission, are applied to
decouple the information of different sources without decoding.
During the second phase, symbols of different sources are forwarded
to the destination in a time division multi-access (TDMA) fashion.
At the destination, the maximum-likelihood (ML) decoding is
performed source-by-source. The protocol of IC-Relay-TDMA requires
the number of relay antennas no less than the number of sources,
i.e., $R_a\ge J$. Through outage analysis, the achievable diversity
gain of the proposed scheme is shown to be
$\min\{J_a(R_a-J+1),R_aM\}$. When {\small$M\le
J_a\left(1-\frac{J-1}{R_a}\right)$}, the proposed scheme achieves
the maximum interference-free (int-free) diversity gain $R_aM$.
Since concurrent transmission is allowed during the first phase,
compared to full TDMA transmission, the proposed scheme achieves the
same diversity, but with a higher symbol rate.
\end{abstract}
{\bf\em Index Terms:} Multi-access relay network, distributed
space-time coding, interference cancellation, orthogonal designs,
quasi-orthogonal designs, cooperative diversity.
\renewcommand{\baselinestretch}{2}

\section{Introduction}
Node cooperation improves the reliability and the capacity of
wireless networks. Recently, many cooperative schemes have been
proposed \cite{LenemanWornell,DSTC-paper,zz-eg-sc,LaTsWo}, and their
multiplexing and diversity gains are analyzed. Most of the pioneer
works on cooperative networks focus on cooperative relay designs
without multi-user interference by assuming that there is one single
transmission task or orthogonal channels are assigned to different
transmission tasks, e.g.,
\cite{LenemanWornell,DSTC-paper,zz-eg-sc,LaTsWo}. As a general
network has multiple nodes each of which can be a data source or
destination, multi-user transmission is a prominent problem in
network communications.

One model on multi-user cooperative communication is interference
relay networks\cite{MoBoNa05}. Multiple pairs of parallel
communication flows are supported by a common set of relays. Each
source targets at one distinct destination. Two transmission schemes
using relays to resolve interference were proposed. The zero-forcing
(ZF) relaying scheme uses scalar gain factors at relays to null out
interference at undesired destinations\cite{WitRan04, Wit06, Niu07}.
The minimum mean square error (MMSE) relaying scheme also uses
scalar gain factors at relays but to minimize the power of
interference-plus-noise at undesired destinations\cite{BerWit05,
Keyi-ICCASP}. Both relaying schemes require the gain factors
calculated at one centralized node having perfect and globe channel
information, then fed back to the relays. These papers discuss the
multiplexing gain and designs of the optimal scalar gain factors,
but do not provide diversity gain analysis. In addition, for general
multi-user cooperative networks, where communication flows may be
unparallel, these schemes cannot be applied straightforwardly. For
example, for a network in which several sources have independent
information for one single-antenna destination, the ZF and MMSE
relaying cannot resolve information collision at the destination.

In this paper, we consider a \emph{multi-access relay network}
(MARN), in which $J$ sources, each equipped with $J_a$ antennas,
send independent information to one $M$-antenna destination through
one $R_a$-antenna relay. We denote this network as a $(J, J_a, R_a,
M)$ MARN. For MARNs, a straightforward scheme is to use full time
division multi-access (TDMA), where sources are allocated orthogonal
channels for both hops of transmissions. Distributed space-time code
(DSTC)\cite{DSTC-paper,DSTC-mulpaper} is performed at the relay to
gain high diversity performance without any channel state
information (CSI). Such a scheme with full TDMA and DSTC at the
relay is denoted as full-TDMA-DSTC. It achieves the maximum
diversity gain $R_a\min\{J_a, M\}$. Since interference is avoided,
this diversity gain is denoted as \emph{interference-free}
(int-free) diversity and provides a natural upperbound on the
spatial diversity gain for all multi-user transmission schemes in
the MARN. However, the spectrum efficiency of full-TDMA-DSTC is low.
Another intuitive scheme is to allow multi-user concurrent
transmission in both hops. The relay conducts decode-and-forward
(DF) by jointly recovering all sources' symbols. However, the
decodings at the relay and the destination induce high processing
complexity which is exponential in the number of sources. In
\cite{Yilmaz-ICC}, ZF beamformers are used in networks with two
sources to make sources' signals orthogonal at the destination. The
relay uses amplified-and-forward (AF). Nevertheless, the beamformer
coefficients or global channel information need to be fed back to
the sources, which induces a high protocol cost. In \cite{ICC09}, we
proposed a scheme called DSTC-ICRec that does not require CSI
feedback to the relay and sources. The scheme allows concurrent
transmission in both hops and uses multiple destination antennas to
perform interference cancellation (IC)\cite{NaSeCa, AlCa, KaJa}.
However, it trades overall diversity for spectral efficiency and
cannot achieve the int-free diversity\cite{TDMAICRec2010}.

%
%

Based on the above discussion, a new scheme, called IC-Relay-TDMA,
is proposed in this paper to allow multi-user concurrent
transmission in the source-relay link. The multi-user interference
is canceled at the multi-antenna relay by the linear IC techniques
proposed in \cite{NaSeCa, AlCa, KaJa}. Then, space-time block code
(STBC) and TDMA are used for the relay to forward signals of
different sources to the destination. The merits of this scheme is
summarized as follows:
\begin{enumerate}
  \item The IC-Relay-TDMA scheme applies not only to MARNs but also to general multi-user cooperative networks
  with multiple destinations and arbitrary patterns of communication
  flows as long as $R_a\ge J$. The scheme requires CSI to be available at the
  receiving nodes only and no feedback is needed. The relay processing is linear and the decoding complexity at the destination
  is linear in the number of sources.
  \item It is proved rigorously that IC-Relay-TDMA achieves a diversity of $\min\{J_a(R_a-J+1),
  R_aM\}$ in a $(J,J_a,R_a,M)$ MARN.
  \item When $M\le
J_a\left(1-\frac{J-1}{R_a}\right)$, IC-Relay-TDMA achieves the
int-free diversity, which is the maximum spatial diversity
achievable for $(J, J_a, R_a, M)$ MARNs. Thus, the concurrent
first-step transmission of the scheme induces no diversity penalty
while improves the spectrum efficiency. The symbol rate of
IC-Relay-TDMA is $\frac{R_o}{J+R_o}$ symbols/user/channel use where
$R_o$ denotes the symbol rate of the STBC used in the
relay-destination link. Since the symbol rate of full-TDMA-DSTC is
$\frac{R_o}{J(1+R_o)}$, IC-Relay-TDMA achieves the same diversity,
but with higher symbol rate, compared to full-TDMA-DSTC.
\end{enumerate}

The rest of the paper is organized as follows. Section
\ref{sec-model} provides the network model. Section
\ref{sec-ICRelay-TDMA} introduces the IC-Relay-TDMA scheme. Its
achievable diversity and symbol rate are discussed in Section
\ref{sec-perana}. Section \ref{sec-Simulation} shows numerical
results and conclusions are given in Section \ref{sec-Conclusion}.
Involved proofs are presented in appendices.

Notation: For a matrix $\mb{A}$, let $\mb{A}^t$, $\mb{A}^*$, and
$\overline{\mb{A}}$ be the transpose, Hermitian, and conjugate of
$\mb{A}$, respectively. $\|\mb{A}\|$ is the Frobenius norm of
$\mb{A}$. $\tr\{\mb{A}\}$ calculates the trace of $\mb{A}$.
$\otimes$ denotes Kronecker product. $\mb{I}_n$ is the $n\times n$
identity matrix. $\mb{0}_n$ is the $n\times n$ matrix of all zeros.
For two matrices of the same dimension, $\mb{A}\succ\mb{B}$ means
that $\mb{A}-\mb{B}$ is positive definite. $f(x)=o(x)$ means
$\underset{x\rightarrow 0^+}{\lim}\frac{f(x)}{x}=0$. $\Exp [x]$
denotes the expected value of the random variable $x$.

\section{Network Models}
\label{sec-model} In this section, we describe two network models
that are used in the paper. First, we introduce the MARN, then the
general multi-user cooperative network.

\begin{figure}[!h]
\centering
\includegraphics[width=4.5in]{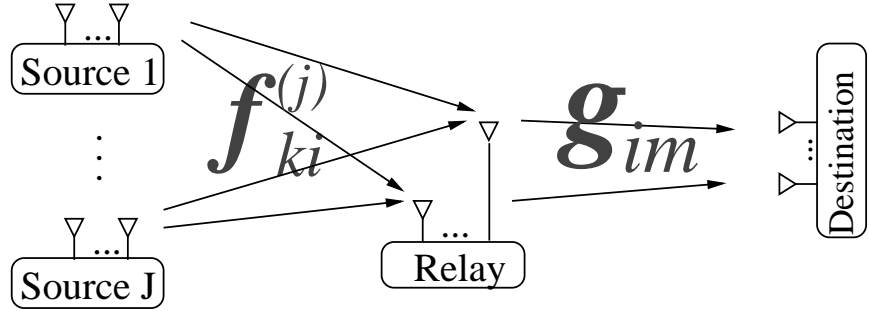}
\caption{Multi-access relay networks.} \label{fig-IC-network}
\end{figure}

Consider a relay network with $J$ sources each with $J_a$ antennas,
one relay with $R_a$ antennas, and one destination with $M$
antennas. There is no direct connection from sources to the
destination, because the sources are far from the destination. The
system diagram is shown in Fig.~\ref{fig-IC-network}.

Denote the fading coefficient from Antenna $k$ ($k=1,\ldots, J_a$)
of Source $j$ ($j=1,\ldots, J$) to Antenna $i$ ($i=1,\ldots, R_a$)
of the relay as $f^{(j)}_{ki}$. The $J_a\times 1$ channel vector
from Source $j$ to relay Antenna $i$ is denoted as $\bff_{i}^{(j)}$
whose $k$-th entry is $f^{(j)}_{ki}$. Denote the fading coefficient
from relay Antenna $i$ to Antenna $m$ of the destination
($m=1,\ldots, M$) as $g_{im}$.  The $R_a \times 1$ channel vector
from the relay to destination Antenna $m$ is denoted as $\bg_{m}$
whose $i$-th entry is $g_{im}$.
All fading coefficients are assumed to be identically and
independently distributed (i.i.d.) with $\mc{CN}(0,1)$ distribution.
We assume a block-fading model with coherence interval $T$.

To allow IC at the relay, we assume that $R_a \ge J$. This can be
realized through user admission control in the upper layers. We
assume that sources have equal numbers of antennas. Our proposed
protocol can be extended straightforwardly to networks where sources
have unequal number of antennas. Further, to focus on the diversity
performance, all sources and the relay are assumed to have the same
average power constraint $P$. The extension to nonuniform power
constraint is also straightforward. Throughout the paper, we assume
global CSI at the destination; but the relay has only the backward
CSI, i.e., channel information from sources to the relay. The
channel information can be obtained by sending pilot sequences from
sources and the relay \cite{DSTC-mulpaper,SunJing2010}. No feedback
or channel estimation forwarding is required. Perfect
synchronization at the symbol level is assumed for the network.


For complexity considerations, two linear constraints are imposed on
the network. For one, the relay linearly transforms its received
signal to generate its output signal without decoding. For the
other, the decoding complexity at the destination is linear in the
number of sources. It can be verified that the full-TDMA-DSTC scheme
mentioned in the introduction section and the DSTC-ICRec scheme
proposed in \cite{ICC09} satisfy these two linear constraints.

\begin{figure}[!h]
\centering
\includegraphics[width=4.5in]{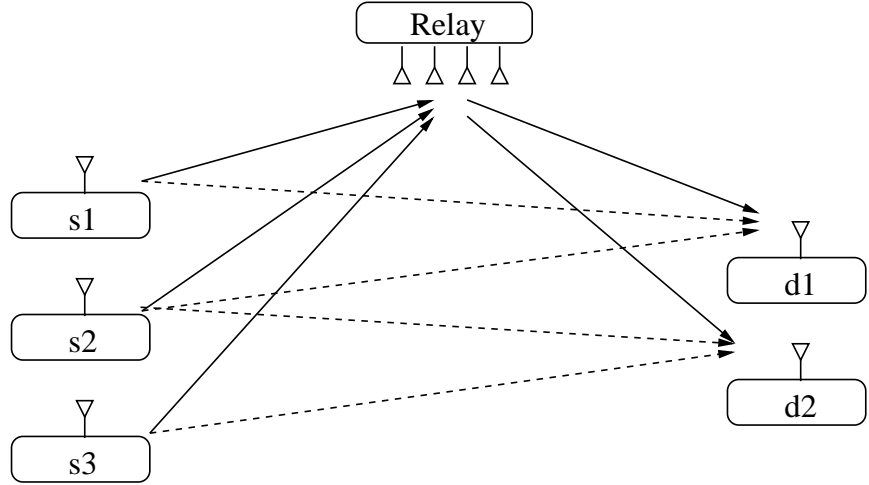}
\caption{General Multi-user cooperative networks. The dash line
denotes communication flows and the solid lines denote physical
links.}\label{fig-genmultiuser}
\end{figure}

The general multi-user cooperative network has $J+N+1$ nodes. $J$
multi-antenna sources, denoted as $S=\left(s_1, s_2, \ldots,
s_J\right)$, send independent information to $N$ multi-antenna
destinations, denoted as $D=\left(d_1, d_2, \ldots, d_N\right)$,
through one $R_a$-antenna relay. In an indoor environment, the
mobile stations can be the sources and the destinations, and the
access points connecting through cables can be the relay. The set of
sources from which Destination $n$ receives information is described
as $I_n=\{s, s\in S\}$. The profile of communication flows of the
whole network can be described as $I=\left(I_1, I_2, \ldots,
I_N\right)$. For example, a network with three single-antenna
sources $S=\left(s_1, s_2, s_3\right)$, two single-antenna
destinations $D=\left(d_1, d_2\right)$, and one four-antenna relay
is shown in Fig. \ref{fig-genmultiuser}. Destination 1 receives
information from Sources 1 and 2, while Destination 2 receives
information from Sources 2 and 3. The profile of communication flows
can be expressed as $I=\left(\{s_1, s_2\}, \{s_2, s_3\}\right)$.
Specifically, when $J=N$ and $I=\left(\{s_1\},\{s_2\},\cdots,
\{s_N\}\right)$, the network becomes the interference relay network
with parallel communication flows. When $N=1$ and $I=\left(\{s_1,
\cdots, s_J\}\right)$, the network becomes the MARN. All nodes are
assumed to be synchronized. Extension to asynchronous networks is
straightforward using the random access and IC methods in
\cite{BHY10}. Although for the clarity of presentation, we present
our protocol using the MARN model, we will show that it can be
applied straightforwardly to this general network model.

\section{IC at the Relay: IC-Relay-TDMA}\label{sec-ICRelay-TDMA}
It is well known that for cooperative networks relaying can improve
communication reliability and coverage. In this paper, we show a new
dimension in the design of multi-user relay networks: IC at relays.
In our MARNs, to improve the spectral efficiency, we allow
concurrent multi-user transmission in the link between the source
and the relay. Since this source-relay link is a multi-antenna
multi-access channel (MAC), the multi-antenna relay has the
potential to cancel the induced multi-user interference. Cancelling
interference at the relay improves the signal to
interference-plus-noise ratio of the relay-destination link and
simplify the signal processing at the destination. Thus, this idea
has the potential of providing good performance when the
relay-destination link is the bottleneck of the network. Based on
the above considerations, we propose a protocol called
IC-Relay-TDMA, in which the relay conducts IC using linear
transformations but not decoding before forwarding int-free signal
to the destination by TDMA. In Subsection \ref{subsec-protocol}, we
describe the protocol for general $(J, J_a, R_a, M)$ MARNs. Its
application in a general multi-user cooperative network is discussed
in Subsection \ref{subsec-general}. Then, the use of IC-Relay-TDMA
in one simple network is illustrated in Subsection
\ref{subsec-example} as an example.

\subsection{The Protocol of IC-Relay-TDMA}\label{subsec-protocol}

\begin{figure}[!h]
\centering
\includegraphics[width=5in]{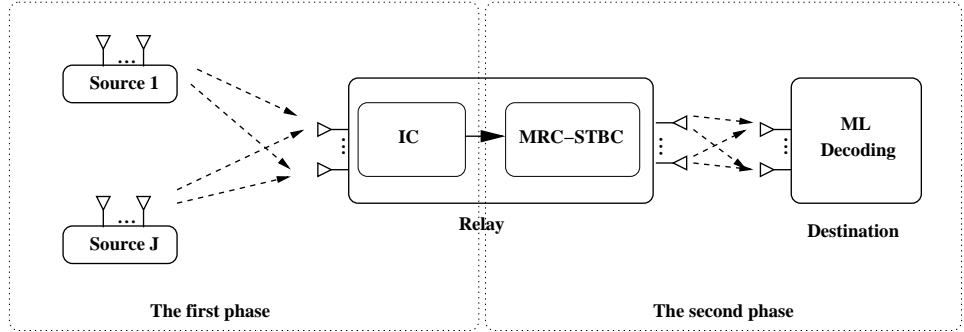}
\caption{Block diagram for IC-Relay-TDMA.} \label{fig-block}
\end{figure}

In this subsection, we explain the protocol of IC-Relay-TDMA. The
protocol consists of two phases. During the first phase, all sources
send information to the relay simultaneously using STBC with ABBA
structure\cite{TBH00,QOD}. The relay overhears superimposed signals
of all source information and conducts multi-user IC\cite{KaJa}.
During the second phase, the relay conducts a scheme called MRC-STBC
to enhance the communication reliability in the second transmission,
then forwards information of different sources in TDMA to the
destination. The destination decodes each source's information
independently. A block diagram of IC-Relay-TDMA is shown in
Fig.~\ref{fig-block}. The details of the protocol and corresponding
formulas are described in the following. First, we consider the
scenarios when $J_a$ is a power-of-2, then extend to the cases that
$J_a$ is not a power-of-2.


\subsubsection*{Phase 1} When the number of antennas at each source is
a power-of-2, i.e., $J_a=2^n, n\in \mc{N}$, each transmitter
constructs $\frac{J_a}{2}$ constellations (e.g. PSK, QAM
constellation and their rotations), denoted as $\mc{S}_u,
u=1,\ldots, \frac{J_a}{2}$. The average power of the constellations
is normalized to be one. The constellations need to satisfy the
following condition for diversity gain:
\be\label{eq-conscon}\underset{u=1:\frac{J_a}{2}}{\sum}c_us_u\neq 0,
\
 \forall s_u\in \mc{S}_u, c_u\in\{-1,0,1\}.\ee
One approach to construct such constellations is through
rotation\cite{ShPa03, SuXia04}. For example, when $J_a=4$, i.e.,
$n=2$, two BPSK constellations can be constructed as
$\mc{S}_1=\{-1,1\}$ and $\mc{S}_2=\{\mr{j},-\mr{j}\}$, where
$\mc{S}_2$ is rotated from $\mc{S}_1$ by $\pi/2$. It can be verified
that $\mc{S}_1$ and $\mc{S}_2$ satisfy the condition in
\eqref{eq-conscon}.

Source $j$ independently and uniformly collects $J_a$ symbols
$s_k^{(j)}, k=1,\ldots, J_a$ from these $J_a/2$ constellations with
$s_{2u-1}^{(j)}, s_{2u}^{(j)}\in \mc{S}_u$. Then, a $J_a\times J_a$
STBC with ABBA structure \cite{QOD,TBH00} is constructed by
\[\mb{S}^{(j)}=\mb{S}_n\left(s^{(j)}_1, s^{(j)}_2,\ldots, s^{(j)}_{J_a}\right),\]
where the function $\mb{S}_n$ maps $J_a$ variables to a $J_a\times
J_a$ matrix through an iterative procedure as \be\nonumber
\mb{S}_n\left(s_1, s_2,\ldots,
s_{J_a}\right)=\left[\begin{array}{cc}\mb{S}_{n-1}\left(s_1,\ldots,
s_{\frac{J_a}{2}}\right)&\mb{S}_{n-1}\left(s_{\frac{J_a}{2}+1},\ldots, s_{J_a}\right)\\
\mb{S}_{n-1}\left(s_{\frac{J_a}{2}+1},\ldots, s_{J_a}\right)&
\mb{S}_{n-1}\left(s_1,\ldots,
s_{\frac{J_a}{2}}\right)\end{array}\right],\ee with $\mb{S}_1(s_1,
s_2)$ an $2\times 2$ Alamouti code. All sources transmit
simultaneously in this phase. The length of this phase is $T_1=J_a$
time slots. It is thus assumed that the coherence interval $T$ is no
less than $J_a$.

Relay Antenna $i$ overhears a $T_1\times 1$ signal vector as
\be\label{eq-1step}\br_{i}=\sum_{j=1}^J
\sqrt{\frac{P}{J_a}}{\mb{S}}^{(j)}\bff_{i}^{(j)}+\bv_{i},\ee where
$\mb{v}_i$ denotes the $J_a \times 1$ additive white Gaussian noise
(AWGN) vector, whose $\tau$-th entry $v_{\tau i}$ is i.i.d.
$\mc{CN}(0,1)$ distributed. Note that the first phase transmission
is virtually a multi-antenna multi-access channel. When $R_a\ge J$,
the IC scheme originally proposed for direction transmission
\cite{KaJa} can be conducted at the relay to fully cancel the
multi-user interference. In \cite{KaJa}, the IC procedure was
discussed explicitly only for a system with at most four antennas at
each source. Here, we describe this procedure for a general system
with $J_a=2^n$ antennas at each source. Without loss of generality,
we discuss how the relay cancels interference from Source 2 to
Source $J$ and obtains an int-free observation of the information of
Source 1.

The IC procedure has two steps. In the first step, the relay
separates the system that communicates $J_a=2^n$ symbols for each
source into $2^{n-1}$ equivalent Alamouti systems. In the second
step, the IC scheme in \cite{AlCa} is applied to each Alamouti
system to iteratively cancel interference from Source 2 to Source
$J$. Denote $\mb{h}_l$ as the $l$-th row of the $2^{n-1}\times
2^{n-1}$ Hadamard matrix $\mb{H}_{n-1}$. Let
{\small$\mb{s}^{(j)}_\o=\left[s^{(j)}_1,s^{(j)}_3,\ldots,s^{(j)}_{2^n-1}\right]^t,\mb{s}^{(j)}_\e=\left[s^{(j)}_2,s^{(j)}_4,\ldots,s^{(j)}_{2^n}\right]^t,{\mb{f}}^{(j)}_{i\o}=\left[f^{(j)}_{1i},f^{(j)}_{3i},\ldots,
f^{(j)}_{(2^{n}-1)i}\right]^t,
{\mb{f}}^{(j)}_{i\e}=\left[f^{(j)}_{2i},f^{(j)}_{4i},\ldots,
f^{(j)}_{2^{n}i}\right]^t, \mb{v}_{i\o}=\left[v_{1i},v_{3i},\ldots,
v_{(2^n-1)i}\right]^t$, and $\mb{v}_{i\e}=\left[v_{2i},
v_{4i},\ldots, v_{(2^n)i}\right]^t$}. As the first step, relay
Antenna $i$ calculates $\tilde{\mb{r}}_{li}=(\mb{h}_l\otimes
\mb{I}_2)\mb{r}_i$ to obtain equivalent Alamouti systems as follows,
\be\label{eq-alasep} \tilde{\mb{r}}_{li}=(\mb{h}_l\otimes
\mb{I}_2)\mb{r}_i=\sum_{j=1:J}
\sqrt{\frac{P}{J_a}}\mb{S}_1\left(\mb{h}_l\mb{s}^{(j)}_\o,\mb{h}_l\mb{s}^{(j)}_\e\right)\left[\begin{array}{c}\mb{h}_l{\mb{f}}^{(j)}_{i\o}\\\mb{h}_l{\mb{f}}^{(j)}_{i\e}\end{array}\right]+\left[\begin{array}{c}\mb{h}_l\mb{v}_{i\o}\\\mb{h}_l\mb{v}_{i\e}\end{array}\right],\
l=1,\ldots, 2^{n-1}.\ee Denote the first and second entries of
$\tilde{\mb{r}}_{li}$ as $\tilde{r}_{li1}$ and $\tilde{r}_{li2}$,
respectively. Due to the Alamouti structure of
$\mb{S}_1\left(\mb{h}_l\mb{s}^{(j)}_\o,\mb{h}_l\mb{s}^{(j)}_\e\right)$.
Eq. \eqref{eq-alasep} can be equivalently rewritten as
\be\label{eq-alatran}
\underset{\hat{\mb{r}}_{li}}{\underbrace{\left[\begin{array}{c}\tilde{r}_{li1}\\
-\ol{\tilde{r}_{li2}}\end{array}\right]}}=\sum_{j=1:J}
\sqrt{\frac{P}{J_a}}\underset{\mb{F}^{(j)}_{li}}{\underbrace{\mb{S}_1\left(\mb{h}_l{\mb{f}}^{(j)}_{i\o},\mb{h}_l{\mb{f}}^{(j)}_{i\e}\right)}}\underset{\hat{\mb{s}}_l^{(j)}}{\underbrace{\left[\begin{array}{c}\mb{h}_l\mb{s}^{(j)}_\o\\\mb{h}_l\mb{s}^{(j)}_\e\end{array}\right]}}+\underset{\hat{\mb{v}}_{li}}{\underbrace{\left[\begin{array}{c}\mb{h}_l\mb{v}_{i\o}\\-\ol{\mb{h}_l\mb{v}_{i\e}}\end{array}\right]}}.\ee
For the second step, the relay cancels interference for each
Alamouti system in multiple iterations, where interference of one
source is cancelled in each iteration. Stack $\hat{\mb{r}}_{li}$ and
$\mb{F}_{li}^{(j)}$ at different relay antenna as
$\hat{\mb{r}}_l=[\hat{\mb{r}}_{l1}^t,\cdots,\hat{\mb{r}}_{lR_a}^t]^t$
and $
\mb{F}^{(j)}_l=\left[\mb{F}_{l1}^{(j)t},\cdots,\mb{F}_{lR_a}^{(j)t}\right]^t$.
Denote $\mc{F}_l(i)$ as the $2(R_a-i-1)\times 2(R_a-i)$ IC matrix to
cancel Source $J-i$ for System $l$; $\mb{r}_l(i)$ and $\mb{F}_l(i)$
as the remaining $2(R_a-i)\times 1$ signal vector and the remaining
$2(R_a-i)\times 2J$ equivalent channel matrix after cancelling
Source $J-i+1$ for System $l$, respectively. The iterative IC
procedures are as follows:
\begin{itemize}
  \item \textbf{Initialization}: $\mb{F}_l(0)=[\mb{F}_l^{(1)},
\ldots,\mb{F}_l^{(J)}]$, $\mb{r}_l(0)=\hat{\mb{r}}_l$.
  \item \textbf{Iteration}: For $i=0$ to $J-2$,
  \begin{enumerate}
    \item Form the IC matrix $\mc{F}_l(i)$ as \\
    \begin{eqnarray}\label{eq-ICmatrix}
    {\mc{F}_l(i)}=\left[\begin{array}{ccccc}-\frac{2{\mb{F}}_{l,J-i,1}^{(i)*}}{\|{\mb{F}}_{l,J-i,1}^{(i)}\|^2}&\frac{2{\mb{F}}_{l,J-i,2}^{(i)*}}{\|{\mb{F}}_{l,J-i,2}^{(i)}\|^2}&\mb{0}_2&\ldots&\mb{0}_2\\
    -\frac{2{\mb{F}}_{l,J-i,1}^{(i)*}}{\|{\mb{F}}_{l,J-i,1}^{(i)}\|^2}&\mb{0}_2&\frac{2{\mb{F}}_{l,J-i,3}^{(i)*}}{\|{\mb{F}}_{l,J-i,3}^{(i)}\|^2}&\ldots&\mb{0}_2\\
    \vdots&\vdots&\vdots&\ddots&\vdots\\
    -\frac{2{\mb{F}}_{l,J-i,1}^{(i)*}}{\|{\mb{F}}_{l,J-i,1}^{(i)}\|^2}&\mb{0}_2&\mb{0}_2&\ldots&\frac{2{\mb{F}}_{l,J-i,M-i}^{(i)*}}{\|{\mb{F}}_{l,J-i,M-i}^{(i)}\|^2}\end{array}\right]_{2(R_a-i-1)\times 2(R_a-i)}
\end{eqnarray}
    \item Cancel interference of Source $J-i$ by multiplying $\mb{r}_l(i)$ with $\mc{F}_l(i)$. The remained
    equivalent received signal can be calculated as
    $\mb{r}_l(i+1)=\mc{F}_l(i)\mb{r}_l(i)$ and the remained equivalent channel matrix can be calculated as
$\mb{F}_l(i+1)=\mc{F}_l(i)\mb{F}_l(i)$.
  \end{enumerate}
\end{itemize}
The $2\times 2$ matrix $\mb{F}_{l,p,q}^{(i)}$ in \eqref{eq-ICmatrix}
denotes the $(p,q)$th $2\times 2$ submatrix of $\mb{F}_l(i)$. After
$J-1$ iterations, information of Sources $J$ to 2 is cancelled and
the remaining signal vector $\mb{r}_l(J-1)$ only contains
information of Source 1. The overall IC matrix that jointly cancels
Sources 2 to $J$ is
$\mc{F}_l\triangleq\prod_{i=0}^{J-2}\mc{F}_l(i)$. To help the
presentation, let $\hat{\mb{r}}_l^{(1)}=\mb{r}_l(J-1)$. From this
iterative procedure, we have \bea\label{eq-remaindb}
\hat{\mb{r}}_l^{(1)}=\mc{F}_l\hat{\mb{r}}_l=\sqrt{\frac{P}{J_a}}\mc{F}_l\mb{F}_l^{(1)}\hat{\mb{s}}_l^{(1)}+\mc{F}_l\mb{v}_l,
\eea where $\mb{v}_l=[\hat{\mb{v}}_{l1}^t,
\hat{\mb{v}}_{l2}^t,\cdots,\hat{\mb{v}}_{lR_a}^t]^t$. A
$2(R_a-J+1)\times 1$ vector observation of Source 1's information is
carried in $\hat{\mb{r}}_l^{(1)}$. Eq. \eqref{eq-remaindb} implies
that the rows of $\mc{F}_l$ are in the null spaces of the columns of
$\mb{F}_l^{(2)}$ to $\mb{F}_l^{(J)}$, i.e.,
$\mc{F}_l\mb{F}_l^{(j)}=\mb{0}$ for $j=2,\ldots, J$. Thus, this IC
process is an iterative realization of ZF. Different from
conventional ZF which uses pseudo-inverse of the channel matrix, the
IC method needs no channel matrix inversion. Similarly, the relay
can obtain int-free vector observations of other sources'
information.

\subsubsection*{Phase 2}
In this phase, the relay conducts a process called
MRC-STBC\cite{JiMRC}, then forwards information of each source to
the destination in different time slots. The destination decodes
source-by-source and jointly recovers the symbols contained in
$\mb{s}_\o^{(j)}$. Without loss of generality, we only consider how
Source 1's information is processed by the relay and decoded at the
destination. The maximum-ratio combining (MRC) step is first
conducted to maximize the SNR of $\hat{\mb{r}}_l^{(1)}$. The MRC can
be represented as \be\label{eq-MRC}
\tilde{\mb{s}}_l^{(1)}=\frac{2\mb{F}_l^{(1)*}\mc{F}_l^*(\mc{F}_l\mc{F}_l^*)^{-1}\hat{\mb{r}}_l^{(1)}}{\tr
(\mb{F}_l^{(1)*}\mc{F}_l^*(\mc{F}_l\mc{F}_l^*)^{-1}\mc{F}_l\mb{F}_l^{(1)})}=\sqrt{\frac{P}{J_a}}\hat{\mb{s}}_l^{(1)}+\underset{\tilde{\mb{v}}_l^{(1)}}{\underbrace{\frac{2\mb{F}_l^{(1)*}\mc{F}_l^*(\mc{F}_l\mc{F}_l^*)^{-1}\mc{F}_l\mb{v}_l}{\tr(\mb{F}_l^{(1)*}\mc{F}_l^*(\mc{F}_l\mc{F}_l^*)^{-1}\mc{F}_l\mb{F}_l^{(1)})}}}.\ee
The entries of $\tilde{\mb{s}}_l^{(1)}$, the vector after MRC, are
soft estimates of the entries of $\hat{\mb{s}}_l^{(1)}$. From
\eqref{eq-alasep}, \eqref{eq-alatran}, \eqref{eq-remaindb}, and
\eqref{eq-MRC}, the covariance matrix of $\tilde{\mb{v}}_l^{(1)}$
can be calculated as
$\frac{J_a}{\tr\{\mb{F}_l^{(1)*}\mc{F}_l^*(\mc{F}_l\mc{F}_l^*)^{-1}\mc{F}_l\mb{F}_l^{(1)}\}}\mb{I}_2$,
which implies that the two noise elements in
$\tilde{\mb{v}}_l^{(1)}$ are i.i.d.. Also, the noise vectors
$\tilde{\mb{v}}_l^{(1)}$ of different Alamouti systems are
independent, due to the orthogonality of the Hadamard matrix
$\mb{H}_{n-1}$, but not identical. Following the MRC step, to
forward Source 1's information, the relay uses generalized
orthogonal STBCs to encode entries of $\tilde{\mb{s}}_l^{(1)}$. We
especially consider generalized orthogonal STBCs because of its full
diversity and symbol-wise decoding \cite{tar99}. Other designs such
as quasi-orthogonal STBCs \cite{QOD} can also be applied but with
higher decoding complexity. In general, consider using a $T_2\times
R_a$ generalized complex orthogonal design that carries $K$ symbols.
Note that each $\tilde{\mb{s}}_l^{(1)}$ contains information of two
symbols. The relay waits for $K$ symbols from $\lceil K/2 \rceil$
Alamouti systems, denoted as $\tilde{s}_k^{(1)}, k=1,\ldots, K$ (the
subscript $l$ is removed without confusion for conciseness), to
generate the $T_2\times R_a$ codeword as \be\label{eq-STBC}
\left[\begin{array}{ccc}\mb{t}_1&\cdots&\mb{t}_{R_a}\end{array}\right]=c\sum_{k=1:K}\left(\real\{\tilde{s}_k^{(1)}\}\mb{A}_k+\mr{j}\imag\{\tilde{s}_k^{(1)}\}\mb{B}_k\right),
\ee where $\mb{t}_i$ is the $T_2\times 1$ signal vector to be
transmitted at relay Antenna $i$; $\mb{A}_k$ and $\mb{B}_k$ are
$T_2\times R_a$ relay encoding matrices for generalized orthogonal
designs as found in (4.67) in \cite{hj}; and
{\small$c\triangleq\sqrt{\frac{PT_2}{R_aK(\frac{P}{2}+\frac{1}{2R_a-2J+1})}}$}
is the power normalization coefficient at the relay. Since the
processing in \eqref{eq-alasep}, \eqref{eq-remaindb},
\eqref{eq-MRC}, and \eqref{eq-STBC} are linear, the transmitted
signal vectors at the relay are linear in its received signal
vectors $\mb{r}_i$. Assume that the coherence interval $T$ is no
less than $T_2$. The relay concurrently forwards $\mb{t}_i$ on
Antenna $i$. The received $T_2\times 1$ signal vector at destination
Antenna $m$ can be expressed as{\be
\mb{x}_m=\sum_{i=1:R_a}\mb{t}_ig_{im}+\mb{w}_m=c\sqrt{\frac{P}{J_a}}\sum_{k=1:K}\left(\real\{\tilde{s}_k^{(1)}\}\mb{A}_k+\mr{j}\imag\{\tilde{s}_k^{(1)}\}\mb{B}_k\right){\mb{g}}_m+{\mb{u}_m},\ee}
\hspace{-4pt}where $\mb{w}_m$ denotes the $T_2\times 1$ AWGN vector
at destination Antenna $m$; $\mb{u}_m$ denotes the equivalent noise
vector,
$\mb{u}_m=c\underset{{k=1:K}}{\sum}\left(\real\{\tilde{v}_k^{(1)}\}\mb{A}_k+\mr{j}\imag\{\tilde{v}_k^{(1)}\}\mb{B}_k
\right)\hat{\mb{g}}_m+{\mb{w}_m}$ with $\tilde{v}_k^{(1)}$ the
additive noise defined in \eqref{eq-MRC}.

At the destination, a $2T_2M\times 1$ vector is formed by stacking
$\mb{x}_m$ into $\tilde{\mb{x}}=\left[\real\{\mb{x}_1\}^t,
\imag\{\mb{x}_1\}^t, \cdots,\right.$ $\left.\real\{\mb{x}_M\}^t,
\imag\{\mb{x}_M\}^t\right]^t$. After straightforward calculation,
the system equation can be written as {\setlength{\arraycolsep}{2pt}
\be\label{eq-recsep}\tilde{\mb{x}}=c\sqrt{\frac{P}{J_a}}\underset{\tilde{\mb{G}}}{\underbrace{\left[\begin{array}{ccccc}\mathcal{A}_1\tilde{\mb{g}}_1&\mathcal{B}_1\tilde{\mb{g}}_1&\cdots&\mathcal{A}_K\tilde{\mb{g}}_1&\mathcal{B}_K\tilde{\mb{g}}_1\\
\vdots&\vdots&\ddots&\vdots&\vdots\\
\mathcal{A}_1\tilde{\mb{g}}_M&\mathcal{B}_1\tilde{\mb{g}}_M&\cdots&\mathcal{A}_K\tilde{\mb{g}}_M&\mathcal{B}_K\tilde{\mb{g}}_M\end{array}\right]}}\left[\begin{array}{c}\real
\tilde{s}_1^{(1)}\\ \imag \tilde{s}_1^{(1)} \\ \vdots \\ \real \tilde{s}_K^{(1)}\\
\imag
\tilde{s}_K^{(1)}\end{array}\right]+\underset{\hat{\mb{u}}}{\underbrace{c\tilde{\mb{G}}\left[\begin{array}{c}\real
\tilde{v}_1^{(1)}\\ \imag \tilde{v}_1^{(1)} \\ \vdots \\ \real \tilde{v}_K^{(1)}\\
\imag
\tilde{v}_K^{(1)}\end{array}\right]+\left[\begin{array}{c}\real
{\mb{w}_1}\\ \imag {\mb{w}_1} \\ \vdots \\ \real {\mb{w}_M}\\
\imag {\mb{w}_M}\end{array}\right]}},\ee} where
{\small\setlength{\arraycolsep}{1pt}\begin{eqnarray*}
\tilde{\mb{g}}_m=\left[\begin{array}{c}\real\{{\mb{g}}_m\}\\
\imag\{{\mb{g}}_m\}\end{array}\right], \
\mathcal{A}_k=\left[\begin{array}{cc}\real\{\mb{A}_k\}&
-\imag\{\mb{A}_k\}\\ \imag\{\mb{A}_k\}&
\real\{\mb{A}_k\}\end{array}\right],\hspace{1mm}\mathcal{B}_k=\left[\begin{array}{cc}-\imag\{\mb{B}_k\}&
-\real\{\mb{B}_k\}\\ \real\{\mb{B}_k\}&
-\imag\{\mb{B}_k\}\end{array}\right]. \end{eqnarray*}}
\hspace{-3pt}With generalized complex orthogonal designs,
$\tilde{\mb{G}}^t\tilde{\mb{G}}=\|\mb{G}\|^2\mb{I}_{2K}$, where $\mb{G}=[\mb{g}_1,\ldots,\mb{g}_M]$. 
Denote $\mb{a}_k$ as a $1 \times 2K$ vector whose $2k-1$ and $2k$
entries are $1$ and $\mr{j}$, respectively, and all the other
entries are zero. The destination can obtain a soft estimate of
$\tilde{s}_k^{(1)}$ by the following
calculation,\be\label{eq-sym1system}
x_k=\mb{a}_k\frac{\tilde{\mb{G}}^t\tilde{\mb{x}}}{\|\mb{G}\|^2}=
c\sqrt{
\frac{P}{J_a}}\tilde{s}_k^{(1)}+c\tilde{v}_k^{(1)}+\frac{w_k}{\|\mb{G}\|},
\ee where $w_k$ is the equivalent noise with $\mc{CN}(0,1)$
distribution and independent for different $k$. $x_k$ is a soft
estimate to $\tilde{s}_k^{(1)}$, which is a linear superposition of
Source 1's symbols from \eqref{eq-alatran}. Without loss of
generality, we assume that $\tilde{s}_k^{(1)}$ provides a soft
estimate to $\mb{h}_k\mb{s}^{(1)}_\o, k=1,\ldots,2^{n-1}$. Denote
{\small\setlength{\arraycolsep}{1pt}$\mb{x}=\left[\begin{array}{ccc}x_1&\ldots&
x_{2^{n-1}}\end{array}\right]^t,
\tilde{\mb{v}}^{(1)}=\left[\begin{array}{ccc}\tilde{v}_1^{(1)}&\ldots&
\tilde{v}_{2^{n-1}}^{(1)}\end{array}\right]^t$} and
{\small\setlength{\arraycolsep}{1pt}$\mb{w}=\left[\begin{array}{ccc}w_1&\ldots&
w_{2^{n-1}}\end{array}\right]^t$}. From \eqref{eq-sym1system}, we
have \be\label{eq-finaleqsys}\mb{x}=c\sqrt{
\frac{P}{J_a}}\mb{H}_{n-1}\mb{s}^{(1)}_\o+\underset{{\mb{u}}}{\underbrace{c\tilde{\mb{v}}^{(1)}+\frac{\mb{w}}{\|\mb{G}\|}}}.
\ee The destination performs ML decoding to decode $\mb{s}^{(1)}_\o$
based on \eqref{eq-finaleqsys} as
\be\label{eq-ML}\arg\min_{\mb{s}}\left(\mb{x}-c\sqrt{
\frac{P}{J_a}}\mb{H}_{n-1}\mb{s}\right)^*\mb{\Sigma}^{-1}_\mb{u}\left(\mb{x}-c\sqrt{
\frac{P}{J_a}}\mb{H}_{n-1}\mb{s}\right),\ee where
$\mb{\Sigma}_\mb{u}$ is the covariance matrix of the equivalent
noise vector $\mb{u}$. After straightforward calculation, we have
\be\label{eq-noisecov}\mb{\Sigma}_\mb{u}=2^{n-1}c^2\diag\left(\frac{1}{\gamma_1},\frac{1}{\gamma_2},\ldots,
\frac{1}{\gamma_{2^{n-1}}}\right)+\frac{1}{\gamma_g}\mb{I}_{2^{n-1}},\ee
where
$\gamma_l=\frac{1}{2}\tr\{\mb{F}_l^{(1)*}\mc{F}_l^*(\mc{F}_l\mc{F}_l^*)^{-1}\mc{F}_l\mb{F}_l^{(1)}\}$
and $\gamma_g=\|\mb{G}\|^2$. Similarly, $\mb{s}_\e^{(1)}$ can be
jointly decoded. Transmission of other sources' symbols can be
performed similarly using orthogonal time slots. To decode all
symbols from all $J$ sources, the destination only needs to conduct
$2J$ procedures of ML decoding of $2^{n-1}$ symbols. The complexity
is linear in the number of sources.

When $J_a$ is not a power-of-2, $2^n\times 2^n$ quasi-orthogonal
STBCs with ABBA structure are used with $2^n$ the smallest
power-of-2 number greater than $J_a$. During the first phase, each
source concurrently transmits the first $J_a$ columns of the block
codes in $T_1=2^n$ time slots. Similar to the case when $J_a$ is a
power-of-2, the resulting multi-user interference can be cancelled
using Eq. \eqref{eq-alasep}, \eqref{eq-alatran}, and
\eqref{eq-remaindb} at the relay by treating $f^{(j)}_{ki}=0$ for
$k=J_a+1, \ldots,2^n$. During the second step, symbols of different
sources are forwarded by MRC-STBC in TDMA. Symbols are decoded
source-by-source at the destination.

\subsection{Application in General Multi-User Cooperative
Networks}\label{subsec-general} IC-Relay-TDMA can be applied to the
general multi-user cooperative networks with multiple destinations.
During the first phase, all $J$ sources send information to the
relay concurrently. The relay separates multi-user signals using IC
without decoding. During the second phase, int-free soft estimates
of each source's symbols are encoded using STBC. The relay
broadcasts each source's block codes using TDMA. All destinations
receive int-free signals from all sources. The destination decodes
its desired information and discards undesired information. Note
that transmission and processing at the relay do not depend on the
relay-destination link and the number of destinations. IC-Relay-TDMA
is robust to the dynamics of the destinations and no
relay-destination channel information is required at the relay. On
the contrary, for ZF relaying and MMSE relaying, the relay needs to
acquire the channel information of new destinations and updates the
scalar gain factors, which takes substantial protocol overhead.
IC-Relay-TDMA can be applied to any patterns of communication flows
when $R_a\ge J$, but ZF relaying and MMSE relaying require the flows
to be parallel. It should be mentioned that IC-Relay-TDMA has a
lower symbol rate than that of ZF and MMSE relaying. For the same
bit rate, larger constellations are required.

\subsection{An Example: IC-Relay-TDMA for a $(2,2,2,1)$ MARN}\label{subsec-example}
In this subsection, we present one example of using IC-Relay-TDMA in
a $(2,2,2,1)$ MARN, where there are two double-antenna sources, one
double-antenna relay, and one single-antenna destination. The
description of the proposed scheme in the previous subsection is
lengthy as it is for a general MARN setting. In this network
example, we will see that some processing are naturally simplified
or become unnecessary and the main ideas behind the scheme are more
clearly illustrated. The complexity at the relay and the destination
can be further reduced.

During the first phase, only one constellation is required and the
constraint on the constellation in \eqref{eq-conscon} becomes
trivial. Both sources collect two symbols from the same
constellation, and concurrently send two Alamouti codes, i.e.,
$\mb{S}^{(j)}=\mb{S}_1\left(s_1^{(j)}, s_2^{(j)}\right),\ j=1,2$.
Since there is one Alamouti system only, the signal separation
illustrated in \eqref{eq-alasep} is also not needed. At the relay,
only one round of IC iteration is needed. The interference of Source
2 can be cancelled by using the IC matrix
{\small$\mc{F}=\left[-\frac{2\mb{F}_1^{(2)*}}{\|\mb{F}_1^{(2)}\|^2},
\frac{2\mb{F}_2^{(2)*}}{\|\mb{F}_2^{(2)}\|^2}\right]$}, where
$\mb{F}_i^{(j)}$ is the Alamouti channel matrix from Source $j$ to
relay Antenna $i$, i.e.,
$\mb{F}_i^{(j)}=\mb{S}_1\left(f_{1i}^{(j)},f_{2i}^{(j)}\right)$. A
$2\times 1$ vector observation of Source 1's symbols can be obtained
from \eqref{eq-remaindb}. During the second phase, after the MRC
represented in \eqref{eq-MRC}, the vector containing information of
each source is encoded into an Alamouti block code and forwarded to
the destination in TDMA. At the destination, the equivalent system
equation for Source 1 can be written as
\be\label{eq-eqsys2}\underset{\hat{\mb{x}}}{\underbrace{\left[\begin{array}{c}x_1\\-\ol{x_2}\end{array}\right]}}=c\sqrt{\frac{P}{2}}\mb{S}_1\left(g_1,
g_2\right)\underset{\mb{s}^{(1)}}{\underbrace{\left[\begin{array}{c}s_1^{(1)}\\s_2^{(1)}\end{array}\right]}}+c\underset{\mb{u}}{\underbrace{\mb{S}_1\left(g_1,
g_2\right)\tilde{\mb{v}}^{(1)}+\tilde{\mb{w}}}},\ee where $x_\tau$,
$\tilde{\mb{v}}^{(1)}$, and $\tilde{\mb{w}}$ denote the received
signal at time slot $\tau$, the equivalent noise vector of the relay
in \eqref{eq-MRC}, and the $2\times 1$ equivalent noise vector at
the destination, respectively. The covariance matrices of
$\tilde{\mb{v}}^{(1)}$ and $\tilde{\mb{w}}$ are $\gamma_f\mb{I}_2$
and $\mb{I}_2$, respectively, where
$\gamma_f=\frac{1}{2}\tr\{\mb{F}^{(1)*}\mc{F}^*\left(\mc{F}\mc{F}^*\right)^{-1}\mc{F}\mb{F}^{(1)}\}$.
Again, for this simple network, the processing in \eqref{eq-recsep}
is not needed. The destination directly performs the ML decoding
based on \eqref{eq-eqsys2}, which, for this network, simplifies to
\be\label{eq-MLdouble}\arg\min_{\mb{s}^{(1)}}\left(\hat{\mb{x}}-c\sqrt{\frac{P}{2}}\mb{S}_1\left(g_1,
g_2\right)\mb{s}^{(1)}\right)^*\mb{\Sigma}_\mb{u}^{-1}\left(\hat{\mb{x}}-c\sqrt{\frac{P}{2}}\mb{S}_1\left(g_1,
g_2\right)\mb{s}^{(1)}\right),\ee where $\mb{\Sigma}_\mb{u}$ is the
covariance matrix of the equivalent noise $\mb{u}$ and after
straightforward calculation,
$\mb{\Sigma}_\mb{u}=\left(c^2\gamma_f\left(\left|g_1\right|^2+\left|g_2\right|^2\right)+1\right)\mb{I}_2$.
Since $\mb{\Sigma}_\mb{u}$ is a multiple of identity independent of
the information symbol, it can be omitted in the ML decoding without
any performance loss. Due to the Alamouti structure in
$\mb{S}_1\left(g_1, g_2\right)$, \eqref{eq-MLdouble} can be further
decomposed into two symbol-wise decodings.

\section{Performance Analysis}\label{sec-perana}
In this section, we provide diversity analysis for the protocol of
IC-Relay-TDMA and discuss its properties. Subsection
\ref{subsec-divana} is on the diversity analysis. In Subsection
\ref{subsec-dis}, we discuss the symbol rate of the scheme and when
it achieves the int-free diversity.

\subsection{Diversity Analysis}\label{subsec-divana}
The diversity of a communication system is defined as the negative
of the asymptotical slope of the bit error rate (BER), $
d=-\underset{P\rightarrow \infty}{\lim}\frac{\log P_b}{\log P}.$ For
fixed constellations, $P_b$ can be replaced with pairwise symbol
error rate (SER). Since the ML decoding in \eqref{eq-ML} is
source-by-source and the network parameters and processing at the
relay and the destination are statistically homogenous, the
diversity of each source is identical. We only need to analyze the
diversity of one source, without loss of generality, Source 1. The
concatenation of two hops of transmission and relay processing make
the calculation of SER extremely difficult. Thus, to aid the
diversity gain analysis, in the following lemma, we provide a method
to calculate the diversity based on a formula with the outage
probability structure without explicitly calculating the SER.

{\renewcommand{\baselinestretch}{1.6}
\begin{lemma}\label{lemma-def}
Define the \emph{instantaneous normalized receive SNR} as
$\gamma=\tr\{\mb{H}_{n-1}^*\mb{\Sigma}_\mb{u}^{-1}\mb{H}_{n-1}\}$.
When the constellations satisfy \eqref{eq-conscon}, the diversity of
the ML decoding in \eqref{eq-ML} can be calculated as
\be\label{eq-div} d=\lim_{\epsilon\rightarrow 0^+}\frac{\log
P(\gamma<\epsilon)}{\log \epsilon}.\ee
\end{lemma}}
\begin{proof}
See the appendix for the proof.
\end{proof}
Lemma \ref{lemma-def} says that diversity can be calculated using
the outage probability of the instantaneous normalized receive SNR.
Thus, diversity can be obtained from the minimum exponent of
$P(\gamma<\epsilon)$. More precisely, a random variable $\gamma$
provides diversity $d$ if
$P(\gamma<\epsilon)=c\epsilon^d+o(\epsilon^d)$ where $c$ is a
constant independent of $\epsilon$. Before the diversity theorem,
the following lemma is introduced.

{\renewcommand{\baselinestretch}{1.6}
\begin{lemma}\label{lemma2}
Let $\gamma_1, \gamma_2, \ldots, \gamma_N, \gamma_g$ be $N+1$
instantaneous normalized receive SNRs. $\gamma_g$ is independent of
$\gamma_n$ for $n=1, 2,\ldots, N$. $\gamma_g$ provides diversity
$d_1$; $\underset{n=1:N}{\sum}\gamma_n$ provides diversity $d_2$. If
$\gamma=\underset{n=1:N}{\sum}\frac{\gamma_n\gamma_g}{\gamma_n+\gamma_g}$,
$\gamma$ provides diversity $\min\{d_1, d_2\}$.
\end{lemma}}
\begin{proof}
It can be shown by straightforward calculation that
$\underset{n=1:N}{\sum}\frac{\gamma_n\gamma_g}{\gamma_n+\gamma_g}<\min\{\underset{n=1:N}{\sum}\gamma_n,
N\gamma_1\}$. The right-hand side has diversity $\min\{d_2, d_1\}$
from Lemma \ref{lemma-def}. Therefore, the diversity of $\gamma$ is
upperbounded by $\min\{d_2, d_1\}$. To show the lowerbounds on the
diversity, the following events are defined:
$\mc{E}\triangleq\{\gamma_1<\gamma_2<\ldots<\gamma_N\}$,
$\mc{E}_n\triangleq \mc{E}\bigcap\{\gamma_{n-1}<\gamma_g<\gamma_n\}$
for $n=1,\ldots, N+1$ where $\gamma_0=0$ and $\gamma_{N+1}=\infty$.
Since $\{\mc{E}_1,\mc{E}_2,\ldots,\mc{E}_{N+1}\}$ is a partition of
$\mc{E}$, we have \begin{eqnarray*}
P(\gamma<\epsilon, \mc{E})&=&P(\gamma<\epsilon, \mc{E}_1)+P(\gamma<\epsilon, \mc{E}_2)+\ldots+P(\gamma<\epsilon, \mc{E}_{N+1})\\
&\le&\left(P(\gamma<\epsilon|\mc{E}_1)+P(\gamma<\epsilon|\mc{E}_2)+\ldots+P(\gamma<\epsilon|\mc{E}_{N+1})\right)\max\{P(\mc{E}_n)\}\\
&\le&\left(NP(\gamma_g<2\epsilon)+P\left(\underset{n=1:N}{\sum}\gamma_n<2\epsilon\right)\right)\max\{P(\mc{E}_n)\}=c\epsilon^{d}+o(\epsilon^d),
\end{eqnarray*}
where $d=\min\{d_1, d_2\}$ and $c$ is a constant independent of
$\epsilon$. For the third inequality, we have used the facts that
$P\left(\gamma<\epsilon|\mc{E}_n\right)<P\left(\frac{\gamma_g\gamma_N}{\gamma_g+\gamma_N}<\epsilon|\mc{E}_n\right)<P(\gamma_g<2\epsilon)$
when $n\le N$, and
$P\left(\gamma<\epsilon|\mc{E}_{N+1}\right)<P\left(\underset{n=1:N}{\sum}\gamma_n<2\epsilon\right)$.
In the third line, the term $\max\{P(\mc{E}_n)\}$ is independent of
$\epsilon$, hence does not affect the diversity. This is true for
any orders of the sequence $\gamma_1, \ldots, \gamma_N$. Note that
$P(\gamma<\epsilon)=\underset{\mc{E}}{\sum}P(\gamma<\epsilon,
\mc{E})$ where the summation is over all possible orders. The
diversity of $\gamma$ is lowerbounded by the minimum of the
exponents of $P(\gamma<\epsilon, \mc{E})$, which is $\min\{d_2,
d_1\}$. Therefore, the diversity is $\min\{d_2, d_1\}$.
\end{proof}

{\renewcommand{\baselinestretch}{1.6}
\begin{theorem}\label{theorem}
In $(J, J_a, R_a, M)$ MARNs, IC-Relay-TDMA achieves a diversity of
$\min\{J_a(R_a-J+1), R_aM\}$ when $R_a\ge J$.
\end{theorem}
}
\begin{proof}
From \eqref{eq-ML} and \eqref{eq-noisecov}, the instantaneous
normalized receive SNR can be calculated as
\begin{eqnarray}
\gamma=\tr\{\mb{H}_{n-1}^*\mb{\Sigma}_\mb{u}^{-1}\mb{H}_{n-1}\}=\underset{l=1:2^{n-1}}{\sum}\left(\frac{2^{n-1}c^2}{\gamma_l}+\frac{1}{\gamma_g}\right)^{-1}.
\end{eqnarray}
Since $\underset{l=1:2^{n-1}}{\sum}{\gamma_l}$ is identical to the
instantaneous normalized receive SNR in a multi-antenna multi-user
system with $J$ $J_a$-antenna users and IC at the $N$-antenna
receiver,
$\frac{1}{2^{n-1}c^2}\underset{l=1:2^{n-1}}{\sum}{\gamma_l}$
provides diversity $J_a(R_a-J+1)$\cite{KaJa-2}. Since $\gamma_g$ is
a Gamma distributed random variable with degree $R_aM$, $\gamma_g$
provides diversity $R_aM$. By Lemma \ref{lemma2}, $\gamma$ has
diversity $\min\{J_a(R_a-J+1), R_aM\}$.
\end{proof}

\subsection{Performance Discussions}\label{subsec-dis}
This subsection discusses the condition for the proposed scheme to
achieve the int-free diversity, the symbol rate, and the complexity
of the proposed scheme. Comparisons with other schemes are also
provided.
\subsubsection*{The int-free diversity condition}
Theorem \ref{theorem} says that IC-Relay-TDMA achieves diversity
$\min\{J_a(R_a-J+1), R_aM\}$. Recall that the int-free diversity is
defined as the maximum achievable diversity for $(J, J_a, R_a, M)$
MARNs without interference, which is $R_a\min\{J_a,M\}$. When
\be\label{eq-fulldiv} M\le J_a\left(1-\frac{J-1}{R_a}\right),\ee
IC-Relay-TDMA achieves diversity $R_aM$, which is equal to the
int-free diversity under \eqref{eq-fulldiv}. Eq. \eqref{eq-fulldiv}
is then called \emph{the int-free diversity condition}. This
condition implies that $M<J_a$, i.e., there are more independent
paths in the source-relay link than the relay-destination link. For
these networks, the bottleneck of transmission is the
relay-destination link. Intuitively, when the source-relay link has
extra degrees of freedom, they can be used for IC without degrading
the total diversity. This is the basic idea behind IC-Relay-TDMA.
Some examples of networks achieving the int-free diversity are $(2,
2, 2, 1)$; $(2, 4, 2, 1)$; and $(2, 2, 4, 2)$ MARNs. To the best of
our knowledge, in multi-antenna MAC, there is no IC method that
achieves full single-user diversity. For MARNs, this is possible due
to the extra relaying step. For networks satisfying
\eqref{eq-fulldiv}, the source-relay link provides enough extra
degrees of freedom to cancel interference at the relay.

\subsubsection*{The symbol rate} During the first phase, each source sends $T_1=2^n$
symbols during $T_1$ time slots. During the second phase, assume
that the relay uses generalized orthogonal STBCs of dimension
$T_2\times R_a$ to carry $K$ symbols. Then, the total number of time
slots in the second phase is $\frac{JT_1T_2}{K}$. Let the symbol
rate of the STBC code used in the second hop as
$R_o\triangleq\frac{K}{T_2}$. Thus, the symbol rate of each source
is $R=T_1/\left(T_1+\frac{JT_1T_2}{K}\right)=\frac{R_o}{J+R_o}$. If
rate-1 codes (e.g., Alamouti code) are used in the second
transmission phase, we have $R_o=1$ and the symbol rate of the
scheme is $\frac{1}{1+J}$.

\subsubsection*{Complexity}We discuss the complexities in terms of the number of sources at the relay
and the destination. Note that the relay needs to cancel the
interfering signals from $J-1$ sources to obtain the int-free signal
from one source and there are $J$ sources needed to be decoupled.
The complexity of the IC is quadratic in the number of sources at
the relay. At the destination, the complexity of ML decoding is
linear in the number sources. Therefore, the relay has higher order
of complexity than the destination in terms of the number of
sources.

\subsubsection*{Comparison with other schemes}
We now compare IC-Relay-TDMA with other schemes in MARNs. Recall
that the proposed IC-Relay-TDMA scheme has concurrent transmission
in the source-relay link only. We first compare it with
full-TDMA-DSTC, which uses TDMA to avoid multi-user interference in
both hops. The second compared scheme is DSTC-ICRec\cite{ICC09},
which allows multi-user concurrent transmission in both hops and IC
at the destination to decouple signals of different sources.
Finally, we introduce DSTC joint-user ML decoding, which is similar
to DSTC-ICRec excepts that, instead of conducting IC then decoding
each source's messages independently, the destination jointly
decodes all sources' messages. Note that the decoding complexity of
this scheme is exponential in the number of sources. Thus, DSTC
joint-user ML decoding does not satisfy the linear constraint at the
destination mentioned in Section \ref{sec-model}, but the other
three schemes satisfy the linear constraints both at the relay and
the destination. We compare diversity, symbol rates, and other
properties of these schemes in Table \ref{table-comp}. The details
on the diversity results in this table can be found in
\cite{Report}.

For networks satisfying the int-free diversity condition,
IC-Relay-TDMA achieves the same diversity as full-TDMA-DSTC with a
higher transmission rate. This is due to the concurrent transmission
and diversity redundancy in the source-relay link. Though the symbol
rate of DSTC-ICRec is higher than that of IC-Relay-TDMA, a higher
dimension constellation can be used for IC-Relay-TDMA to achieve the
same bit rate with faster decaying error probability. DSTC
joint-user ML decoding achieves the maximum int-free diversity with
a symbol rate higher than that of IC-Relay-TDMA. However, the
decoding complexity of the DSTC joint-user ML decoding is
exponential in the number of sources, which is very demanding when
$J$ is large. This implies that the proposed IC-Relay-TDMA trades
decoding complexity for symbol rate without losing diversity for
networks satisfying the int-free diversity condition. It should also
be noted that IC-Relay-TDMA requires backward CSI at the relay,
while the other three schemes do not require any CSI at the relay.
Backward CSI can be obtained via training and does not need any
feedback.

\section{Numerical Results}\label{sec-Simulation}
In this section, we show simulated BER performance of IC-Relay-TDMA
and its comparison with other schemes. In all figures, the
horizontal axis represents the average transmit SNR, measured in dB.
Since the noises at all nodes are normalized, $P$ is equal to the
average transmit SNR. The vertical axis represents the BER.

\begin{figure}[!h]
\centering
\includegraphics[height=5in, angle=-90]{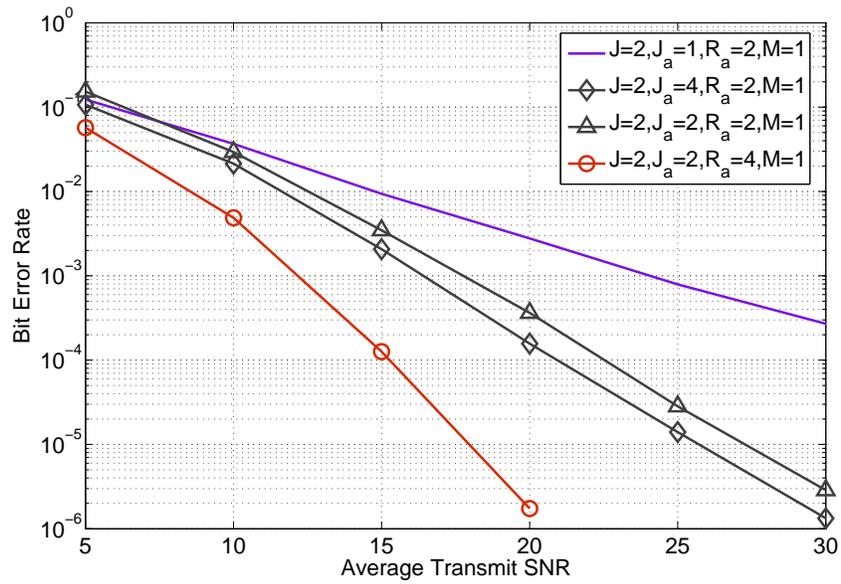}
\caption{BER performance IC-Relay-TDMA, under BPSK modulation.}
\label{fig-IC-Relay-TDMA}
\end{figure}

Fig.~\ref{fig-IC-Relay-TDMA} is on the BERs of IC-Relay-TDMA under
four network scenarios: Network 1: $(2, 1, 2, 1)$ MARN; Network 2:
$(2, 2, 2, 1)$ MARN; Network 3: $(2, 4, 2, 1)$ MARN; and Network 4:
$(2, 2, 4, 1)$ MARN. For the first three networks, Alamouti codes
are used; and for Network 4, the rate $3/4$ generalized orthogonal
STBC shown in (4.103) of \cite{hj} is used. All networks apply BPSK
modulation. Networks 2, 3, and 4 satisfy the int-free diversity.
Fig.~\ref{fig-IC-Relay-TDMA} shows that for these three networks,
the proposed scheme achieves the int-free diversity 2 for Networks 2
and 3; and 4 for Network 4. Network 1 does not satisfy the int-free
condition in \eqref{eq-fulldiv}. Fig.~\ref{fig-IC-Relay-TDMA} shows
that Network 1 has diversity 1, which is less than 2, the int-free
diversity. These simulation results verify Theorem \ref{theorem}.

\begin{figure}[!h]
\centering
\includegraphics[height=5in, angle=-90]{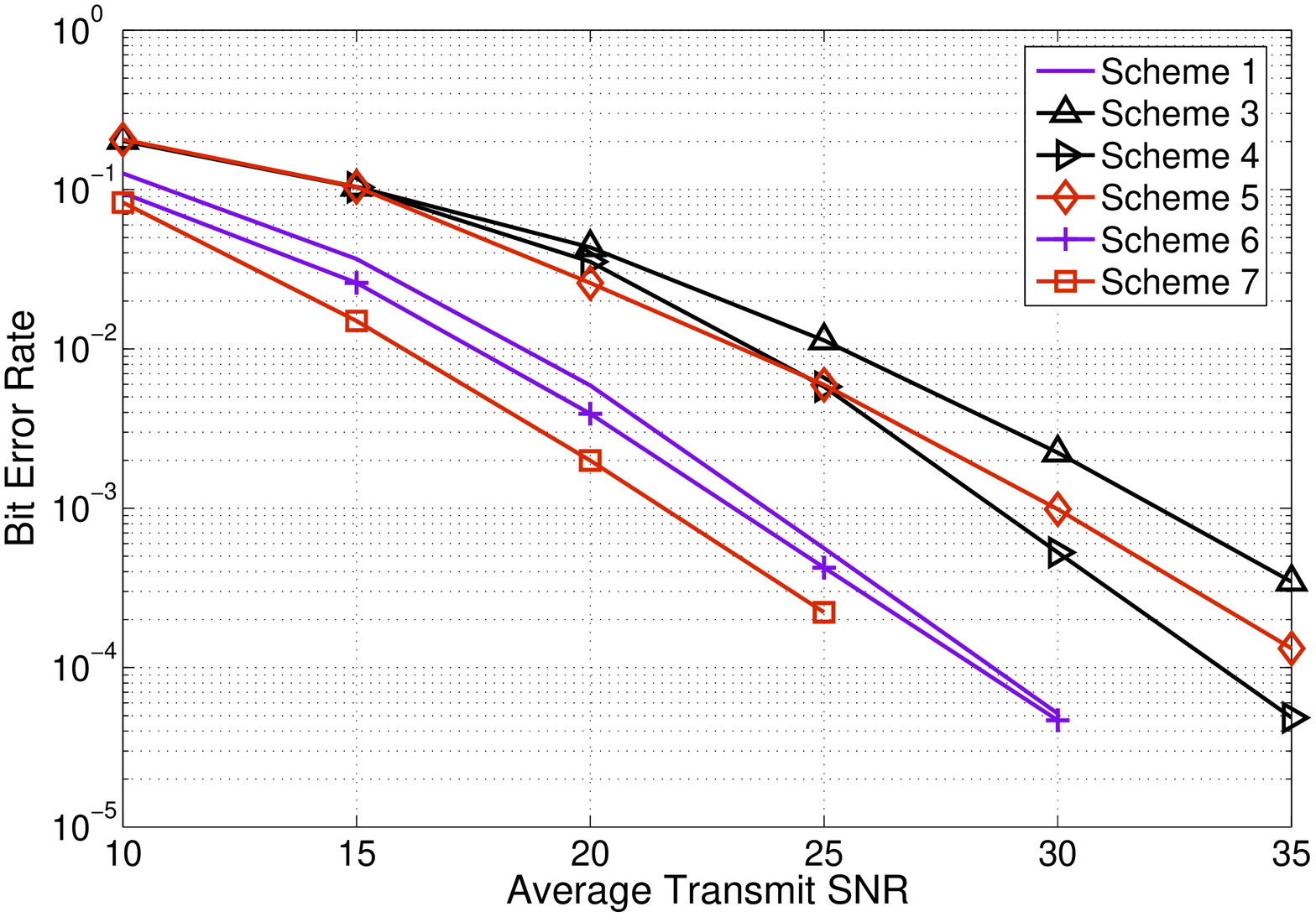}
\caption{Performance comparison in a $(2,2,2,1)$ MARN, under 1
bit/user/channel use.} \label{fig-4div}
\end{figure}

In what follows, we compare the proposed IC-Relay-TDMA (Scheme 1)
with DSTC-ICRec (Scheme 2), full-TDMA-DSTC (Scheme 3),
full-TDMA-DSTC CIR (Scheme 4), DSTC joint-user ML decoding (Scheme
5), IC-Relay-TDMA DF (Scheme 6), joint-DF-TDMA (Scheme 7) in the
$(2,1,2,2)$ and $(2,2,2,1)$ MARNs. Schemes 1, 2, 3, and 5 are
discussed in Subsection \ref{subsec-dis}. To rule out the effect of
the difference in the CSI requirements for Scheme 1 (the relay needs
to know its channels with the transmitters) and Scheme 3 (no channel
information at the relay), Scheme 4, originally proposed in
\cite{DSTC-OD} for single-user relay networks, is included as well.
In this scheme, the relay uses its knowledge of the backward CSI to
equalize the phase shift of the source-relay link, then forwards
information to the destination by Alamouti DSTC. To allow decoding
at the relay, Scheme 6 is introduced, which is similar to Scheme 1
except that all sources' symbols are decoded after IC at the relay
and re-modulated by the same constellation before forwarding. For
Scheme 7, the relay jointly decodes symbols from both sources
without IC before forwarding each source's information using
Alamouti DSTC in TDMA. Note that Schemes 1, 2, 3, 4 satisfy the
linear constraints introduced in Section \ref{sec-model}; whereas
Schemes 5,6,7 do not. For Scheme 5, the decoding complexity is
exponential in $J$ at the destination. The relay's decoding
complexity for Schemes 6 and 7 are linear and exponential in $J$,
respectively. Schemes 1, 4, 6, 7 require backward CSI at the relay,
but the other three schemes need no CSI at the relay. To achieve 1
bit/user/channel use for all schemes, the modulation constellations
used for the schemes are 8PSK, QPSK, 16PSK, 16PSK, QPSK, 8PSK, 8PSK,
respectively. Since the destination in the $(2,2,2,1)$ MARN has only
single-antenna, Scheme 2 is excluded from the comparison in
Fig.~\ref{fig-4div}.

\begin{figure}[!h]
\centering
\includegraphics[height=5in, angle=-90]{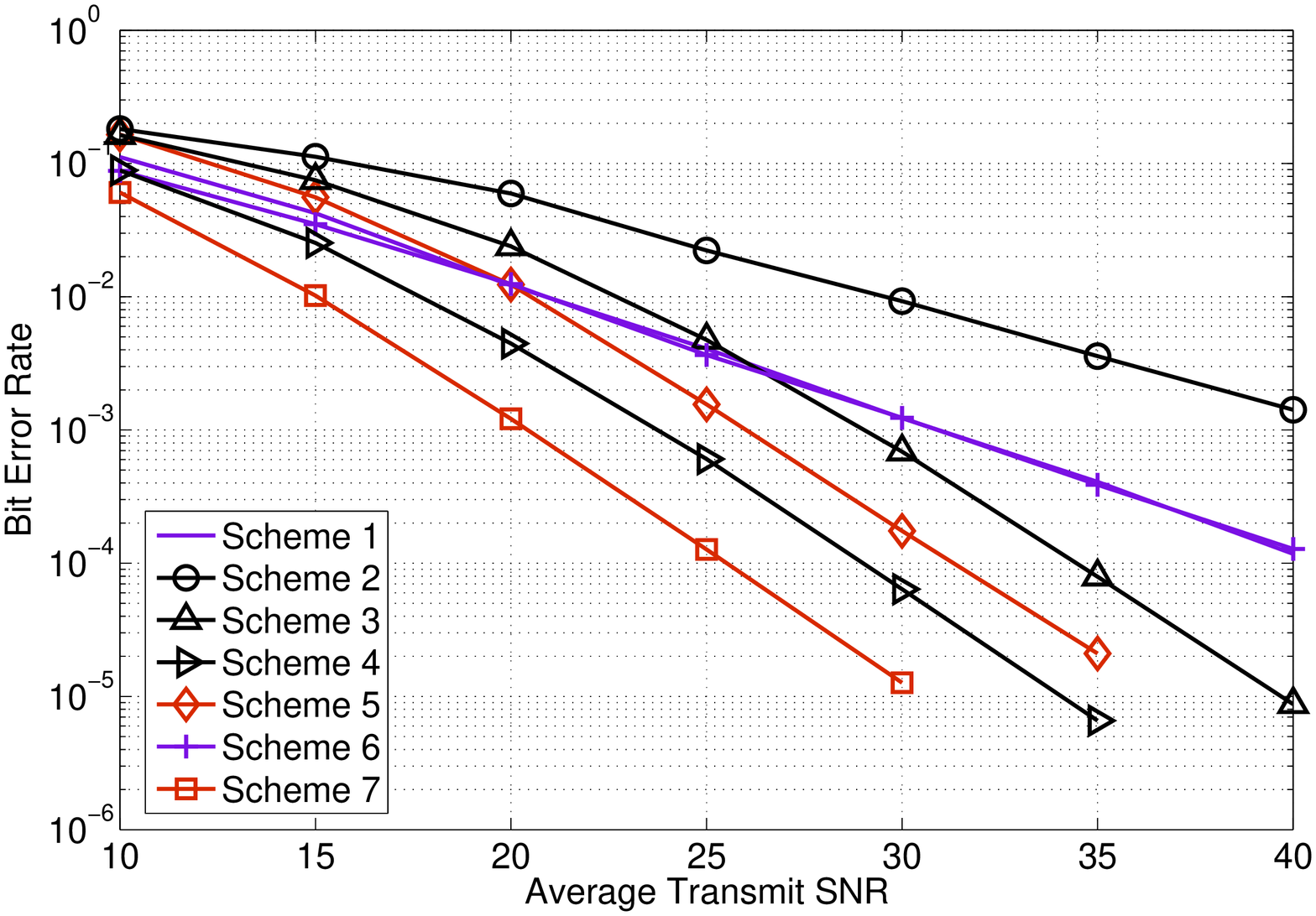}
\caption{Performance comparison in a $(2,1,2,2)$ MARN, under 1
bit/user/channel use.} \label{fig-divloss}
\end{figure}

We first compare the BER of Scheme 1 with the other linear schemes
(Schemes 2, 3, 4). In the $(2,1,2,2)$ MARN (Fig.~\ref{fig-divloss}),
Schemes 3 and 4 achieve a diversity gain of 2, which is the int-free
diversity gain. Schemes 1 and 2 achieve a diversity gain of 1 only.
For this network, since the int-free condition is not satisfied, the
proposed Scheme 1 performs worse than Scheme 4 for the entire
simulated SNR range. Although it is worse than Scheme 3 for SNR
larger than 27 dB due to the diversity loss, it outperforms Scheme 3
when the SNR is smaller than 27 dB due to its higher symbol rate.
Compared with Scheme 2, the proposed scheme is superior for the
simulated SNR range. At the BER level of $10^{-2}$, it is about
10~dB better. For the $(2,2,2,1)$ MARN (Fig.~\ref{fig-4div}), the
int-free condition is satisfied. The figure shows that Schemes 1 and
4 achieve a diversity gain of 2, while the diversity of Scheme 3 is
slightly less than 2. This is because for Scheme 3, there is a $\log
P$ factor in the error rate
formula\renewcommand{\baselinestretch}{1}\footnote{{If
quasi-orthogonal designs are used as the distributed STBC, the $\log
P$ factor does not appear and diversity 2 can be achieved as proved
in \cite{DSTC-OD}. However, the use of quasi-orthogonal designs
requires the coherent interval to be 4. In this simulation, $T=2$
and Alamouti designs are used at both the relay and the
transmitters.}}. As $P$ increases, the diversity should approach 2.
For Schemes 1 and 4, the MRC and equalization eliminate the $\log P$
factor. Scheme 2 cannot be used for this network because the
destination has only one antenna and cannot conduct full IC. The
array gain of Scheme 1 is higher compared to both Schemes 3 and 4,
since a lower-dimension constellation is used to achieve the same
bit rate. At the BER level of $10^{-3}$, it is better than Schemes 3
and 4 by 10~dB and 5~dB, respectively. From the comparison, we can
conclude that Scheme 1 is expected to outperform other linear
schemes for MARNs satisfying the int-free condition, e.g., the
$(2,2,2,1)$ MARN.

Next, we compare Scheme 1 with the schemes not satisfying the linear
constraints (Schemes 5, 6, 7). Scheme 1 is first compared with
Scheme 5. Scheme 5 achieves the int-free diversity from Table
\ref{table-comp}, thus naturally having better BER than Scheme 1 in
the high SNR regime for the $(2,1,2,2)$ MARN
(Fig.~\ref{fig-divloss}). For SNR smaller than $20$~dB, Scheme 1
outperforms Scheme 5. In the $(2,2,2,1)$ MARN (Fig.\ref{fig-4div}),
where both schemes achieve the int-free diversity, Scheme 1
outperforms Scheme 5 in the entire SNR, e.g., it is about 6 dB
better when BER$=10^{-3}$. The gain is obtained because in Scheme 1,
user interference is avoided in the second step and the received
signal quality is high. Then, Scheme 1 is compared with Scheme 6,
which has additional decoding after IC. From both
Fig.~\ref{fig-divloss} and \ref{fig-4div}, we can observe that there
is no diversity improvement by additional decoding after IC for
Scheme 6. For the array gain, Scheme 6 outperforms in the low SNR
regime (about 0.5~dB in Fig.~\ref{fig-divloss} and 1.3~dB in
Fig.~\ref{fig-4div}); and has the same BER as Scheme 1 in the high
SNR regime. This is because the BER performance is mainly restricted
by interference in the high SNR regime. Finally, Scheme 1 is
compared with Scheme 7, which allows joint decoding at the relay. We
can see that joint decoding at the relay helps the network to
achieve the int-free diversity, e.g., 2 in the $(2,1,2,2)$ MARN
(Fig.~\ref{fig-divloss}), in addition to an improved array gain
compared to Scheme 1, e.g., about 10~dB in the $(2,1,2,2)$ MARN
(Fig.~\ref{fig-divloss}) and 2~dB in the $(2,2,2,1)$ MARN
(Fig.~\ref{fig-4div}). However, the relay needs to jointly decode
both user's symbols. Two symbols with 8PSK constellation are jointly
recovered in the $(2,1,2,2)$ MARN, and four symbols with 8PSK
constellation in the $(2,2,2,1)$ MARN. Since the relay does not need
to decode in Scheme 1, the complexity of Scheme 7 is much higher
compared to Scheme 1, especially in the $(2,2,2,1)$ MARN.

\section{Conclusions}\label{sec-Conclusion}
This paper is concerned with multi-user transmission and detection
schemes for multi-access relay networks, in which multiple sources
communicate with one destination by a common relay in two hops. For
complexity considerations, the nodes in the network have two linear
constraints: the relay generates its forward signals by linearly
transforming its received signals; the destination has linear
decoding complexity in the number of sources. A new scheme, called
IC-Relay-TDMA, is proposed to cancel interference at the relay and
forward the int-free observations of sources' information in TDMA to
the destination. IC-Relay-TDMA efficiently allows multi-users to
communicate simultaneously in the first hop to enhance transmission
rate. Through rigorous analysis and simulations, it is shown that
IC-Relay-TDMA achieves diversity $\min\{J_a(R_a-J+1), R_aM\}$. When
the number of destination antennas is no higher than
$J_a(1-\frac{J-1}{R_a})$, the maximum int-free diversity $R_aM$ is
achievable, with a higher symbol rate compared to the full-TDMA-DSTC
scheme.

\section*{Appendix: Proof of Lemma \ref{lemma-def}}
%
First, we show the scenario when the decoding in \eqref{eq-ML} is
symbol-wise, i.e., $n=1$, then the scenario for multiple symbol
joint decoding, i.e., $n\ge 2$.

For $n=1$, $\mb{H}_{n-1}=1$,
${\Sigma}_{\mb{u}}=\frac{c^2}{\gamma_1}+\frac{1}{\gamma_g}$, and
$\mb{s}_\o^{(1)}=s_1^{(1)}$ from \eqref{eq-ML}. The ML decoding is
symbol wise. Let $\alpha=\frac{c^2}{2J_a}$,
$\gamma={\Sigma}_{\mb{u}}^{-1}$, and $s_1^{(1)}-s_1^{(1)'}=\Delta
s_1^{(1)}$. The pairwise SER of decoding $s_1^{(1)}$ to $s_1^{(1)'}$
can be written as \be\label{eq-SERsin} P(s_1^{(1)}\rightarrow
s_1^{(1)'})=\underset{f^{(j)}_{ki},g_{im}}{\Exp}Q\left(\sqrt{\alpha
P\left|\Delta s_1^{(1)}\right|^2\gamma} \right),\ee where $Q(x)$
denotes the Gaussian Q function. Note that
$Q(x)\ge\min\left\{\frac{1}{5}e^{-\frac{x^2}{2}},
\frac{1}{3x}e^{-\frac{x^2}{2}} \right\}$ and both
$\frac{1}{3x}e^{-\frac{x^2}{2}}$ and $\frac{1}{5}e^{-\frac{x^2}{2}}$
are decreasing functions. Thus, for any $\epsilon\ge 0$,
$P(s_1^{(1)} \rightarrow s_1^{(1)'}) \ge P(\gamma\le
\epsilon)\min\left\{\frac{1}{5}e^{- \alpha P \Delta^2
\epsilon},\frac{e^{-\alpha P \Delta^2 \epsilon}}{3\sqrt{P \alpha
\Delta^2 \epsilon}}\right\}$ with $\Delta$ the minimum distance
between any two constellation points. Let $\epsilon =P^{-1}$, we
have {\small$ P(s_1^{(1)}\rightarrow s_1^{(1)'}) \ge P(\gamma \le
\epsilon) \underset{\zeta}{\underbrace{\min\left\{
\frac{1}{5}e^{-\alpha\Delta^2},\frac{e^{-\alpha\Delta^2}}{3\sqrt{\alpha}\Delta}
\right\}}}$}, where $\zeta$ is a constant independent of $\epsilon$.
Thus, diversity can be calculated as
\[ d=-\lim_{P\rightarrow \infty}\frac{\log\ P(s_1^{(1)}\rightarrow s_1^{(1)'})}{\log\ P}\le
\lim_{\epsilon\rightarrow 0^+}\frac{\log\ P(\gamma<\epsilon)+\log\
\zeta}{\log\ \epsilon}=\lim_{\epsilon\rightarrow 0^+}\frac{\log\
P(\gamma<\epsilon)}{\log \epsilon}.\] This shows that the diversity
is upperbounded by the right-hand-side (RHS) of \eqref{eq-div}.
Next, we show that the diversity is also lowerbounded by the RHS of
\eqref{eq-div}. Denote $\beta$ as the maximum distance between any
two constellation points. For any $\epsilon\ge 0$, using the
Chernoff bound on SER and noticing that $e^{-\frac{\beta^2 \alpha P
\gamma}{2}}$ is a decreasing function with $\gamma$, we have
\be\label{eq-divproof} P(s_1^{(1)}\rightarrow s_1^{(1)'})
<\underset{\gamma}{\Exp} e^{-\frac{\beta^2\alpha P
\gamma}{2}}=\int_0^\epsilon e^{-\frac{\beta^2\alpha P
\gamma}{2}}f(\gamma)d\gamma+\int_\epsilon^\infty
e^{-\frac{\beta^2\alpha P
\gamma}{2}}f(\gamma)d\gamma<P(\gamma<\epsilon)+e^{-\frac{\beta^2\alpha
P \epsilon}{2}},\ee where $f(\gamma)$ is the probability density
function of $\gamma$.  Let $\epsilon=P^{-n}, 0<n<1$. As $P$
increases, the RHS of \eqref{eq-divproof} is dominated by
$P(\gamma<\epsilon)$. The diversity can be lowerbounded as $d\ge
\underset{{\epsilon\rightarrow 0^+}}{\lim}\frac{n\log
P(\gamma<\epsilon)}{\log\ \epsilon}$. Since $n$ can be chosen very
close to $1$, the lowerbound and upperbound converge.

Now, we consider the case of $n>1$. Denote $\Delta \mb{s}_\o^{(1)}=
\mb{s}_\o^{(1)}-\mb{s}_\o^{(1)'}$. The pairwise error probability of
decoding a vector of $2^{n-1}$ symbols $\mb{s}_\o^{(1)}$ to
$\mb{s}_\o^{(1)'}$ in the ML decoding in \eqref{eq-ML} can be
written as \be\label{eq-PEP}P\left(\mb{s}_\o^{(1)}\rightarrow
\mb{s}_\o^{(1)'}
\right)=\underset{f_{ki}^{(j)},g_{im}}{\Exp}Q\left(\sqrt{\alpha
P\underset{\Gamma}{\underbrace{\Delta
\mb{s}_\o^{(1)*}\mb{H}_{n-1}^*\mb{\Sigma}_\mb{u}^{-1}\mb{H}_{n-1}\Delta
\mb{s}_\o^{(1)}}}}\right).\ee Since
$\mb{H}_{n-1}^*\mb{\Sigma}_\mb{u}^{-1}\mb{H}_{n-1}\prec
\tr\{\mb{H}_{n-1}^*\mb{\Sigma}_\mb{u}^{-1}\mb{H}_{n-1}\}\mb{I}_{2^{n-1}}$,
we have that
$\Gamma<\tr\{\mb{H}_{n-1}^*\mb{\Sigma}_\mb{u}^{-1}\mb{H}_{n-1}\}\|\Delta
\mb{s}_\o^{(1)}\|^2<2^{n-1}\tr\{\mb{H}_{n-1}^*\mb{\Sigma}_\mb{u}^{-1}\mb{H}_{n-1}\}\beta^2$
with $\beta$ the maximum distance between any two points in all
constellations. Thus, a lowerbound on \eqref{eq-PEP} can be obtained
as $P_l=Q\left(\sqrt{{\alpha
2^{n-1}P}\tr\{\mb{H}_{n-1}^*\mb{\Sigma}_\mb{u}^{-1}\mb{H}_{n-1}\}\beta^2}\right)$.
The diversity of $P_l$ upperbounds that of
$P\left(\mb{s}_\o^{(1)}\rightarrow \mb{s}_\o^{(1)'} \right)$. Recall
that $P_l$ is similar to \eqref{eq-SERsin}. The diversity of the
lowerbound can be evaluated by \eqref{eq-div} using
$\tr\{\mb{H}_{n-1}^*\mb{\Sigma}_\mb{u}^{-1}\mb{H}_{n-1}\}$ as the
instantaneous normalized receive SNR. Thus, the diversity obtained
by using
$\gamma=\tr\{\mb{H}_{n-1}^*\mb{\Sigma}_\mb{u}^{-1}\mb{H}_{n-1}\}$ in
\eqref{eq-div} upperbounds that of
$P\left(\mb{s}_\o^{(1)}\rightarrow \mb{s}_\o^{(1)'} \right)$.

Next, we find the lowerbound on the diversity of
$P\left(\mb{s}_\o^{(1)}\rightarrow \mb{s}_\o^{(1)'} \right)$. Since
the entries of $\mb{s}_\o^{(1)}$ are collected from finite
constellations satisfying \eqref{eq-conscon}, there exists a
positive number $\theta$ that lowerbounds all
$\left|\mb{h}_l\Delta\mb{s}_\o^{(1)}\right|^2$. Noticing that
$\mb{\Sigma}_\mb{u}$ is diagonal from \eqref{eq-noisecov}. We have
$\Gamma=\underset{l=1:2^{n-1}}{\sum}\lambda_l^{-1}\left|\mb{h}_l\Delta
\mb{s}_\o^{(1)}\right|^2>\underset{l=1:2^{n-1}}{\sum}\lambda_l^{-1}\theta^2=\tr\{\mb{H}_{n-1}^*\mb{\Sigma}_\mb{u}^{-1}\mb{H}_{n-1}\}\frac{\theta^2}{2^{n-1}}$
with $\lambda_l$ the $l$-th diagonal entry of $\mb{\Sigma}_\mb{u}$.
 Therefore, the
diversity obtained by using
$\gamma=\tr\{\mb{H}_{n-1}^*\mb{\Sigma}_\mb{u}^{-1}\mb{H}_{n-1}\}$ in
\eqref{eq-div} lowerbounds that of
$P\left(\mb{s}_\o^{(1)}\rightarrow \mb{s}_\o^{(1)'} \right)$. The
diversity of $P\left(\mb{s}_\o^{(1)}\rightarrow \mb{s}_\o^{(1)'}
\right)$ can be calculated by using
$\gamma=\tr\{\mb{H}_{n-1}^*\mb{\Sigma}_\mb{u}^{-1}\mb{H}_{n-1}\}$ in
\eqref{eq-div}.

\renewcommand{\baselinestretch}{1}
\bibliographystyle{IEEEtran}
\bibliography{IEEEabrv,IC-Relay-TDMA}

\newpage
\begin{table}[!h]
  \centering
    \caption{Performance comparison in $(J, J_a, R_a, M)$ MARNs.}\label{table-comp}
  \begin{tabular}{|c|c|c|c|c|c|}
  \hline
  Scheme & Concurrent Transmission & Diversity & Symbol Rate & Linearity \\
  \hline
  IC-Relay-TDMA & only the user-relay link & $\min\{J_a(R_a-J+1), R_aM\}$ & $\frac{R_o}{J+R_o}$ & Yes \\
  \hline
  full-TDMA-DSTC & none & $R_a\min\{J_a, M\}$ & $\frac{R_o}{J(1+R_o)}$ &  Yes \\
  \hline
  DSTC joint-user ML decoding & both links & $R_a\min\{J_a, M\}$ & $\frac{1}{2}$ &  No \\
  \hline
  DSTC-ICRec & both links & \begin{tabular}{c}$\min\{J_a,M-1\}$\\ for $J=2, J_a=1,2, R_a=2$ \end{tabular}&
  $\frac{1}{2}$ & Yes\\
  \hline
\end{tabular}
\end{table}

\end{document}